\documentclass[reprint,final,prl,superscriptaddress,amsfonts,showpacs,letterpaper,pdftex]{revtex4-1}
\usepackage{graphicx,multirow,bm,amsmath}
\usepackage{booktabs}
\usepackage{hyperref}
\usepackage{bbm}
\usepackage[usenames,dvipsnames]{color}
\usepackage[table,xcdraw]{xcolor}

\begin{document}

	\title{High $n$-type thermoelectric power factor and efficiency in Ba$_{2}$BiAu from a highly dispersive band}
	\author{Junsoo Park}
	\email{jsyony37@ucla.edu}
	\affiliation{Department of Materials Science \& Engineering, UCLA, Los Angeles, CA 90095, USA} 
	\affiliation{Department of Applied Physics, Yale University, New Haven, CT 06511, USA} 
	\affiliation{Energy Sciences Institute, Yale University, West Haven, CT 06516, USA} 	
	\author{Yi Xia}
	\email{yimaverickxia@gmail.com}
	\affiliation{Nanoscience and Technology Division, Argonne National Laboratory, Argonne, IL  60439, USA}
	\author{Vidvuds Ozoli\c{n}\v{s}}
	\email{vidvuds.ozolins@yale.edu}
	\affiliation{Department of Applied Physics, Yale University, New Haven, CT 06511, USA} 	
	\affiliation{Energy Sciences Institute, Yale University, West Haven, CT 06516, USA} 	
	\date{\today} 
	\begin{abstract}
Using first-principles density-functional theory calculations, we predict the potential for unprecedented thermoelectric efficiency $zT=5$ at 800 K in $n$-type Ba$_{2}$BiAu full-Heusler compound. Such a high efficiency arises from an intrinsically ultralow lattice thermal conductivity coupled with a very high power factor reaching 7 mW m$^{-1}$ K$^{-2}$ at 500 K. The high power factor originates from a light, sixfold degenerate conduction band pocket along the $\Gamma$-X direction. Weak acoustic phonon scattering and sixfold multiplicity combine to yield high mobility and high Seebeck coefficient. In contrast, the flat-and-dispersive (a.k.a. low-dimensional) valence band of Ba$_{2}$BiAu fail to generate a high power factor due to strong acoustic phonon scattering. The Lorenz numbers at optimal doping are smaller than the Wiedemann-Franz value, an integral feature for $zT$ enhancement as electrons are the majority heat carriers. 
	\end{abstract}
	\maketitle


Thermoelectric materials can be used to directly convert heat into electricity and vice versa, and have technological applications in energy harvesting and refrigeration. The indicator of thermoelectric efficiency is the dimensionless figure of merit $zT= \frac{\alpha^{2}\sigma T}{\kappa}$. Here, $\alpha^{2}\sigma$ is the thermoelectric power factor (PF), composed of the Seebeck coefficient ($\alpha$) and electronic conductivity ($\sigma$). The total thermal conductivity ($\kappa$) is the sum of lattice thermal conductivity ($\kappa_{\text{lat}}$) and thermal conductivity due to electrons ($\kappa_{\text{e}}$). For large-scale deployment, thermoelectric materials with $zT>4$ are highly coveted. Unfortunately, conversion efficiencies of even the best thermoelectrics have been stagnant at $zT<3$, only in rare cases surpassing $zT=2$, and even then only at high temperatures \cite{snsenature,last,pbtes,multipleskut}. At room temperature, only alloys of Bi$_{2}$Te$_{3}$ offer competitive performance \cite{bisbtesuperlattice,bisbtenano,bisbtedislocation,bitese}. 

Clearly, a desirable thermoelectric material would feature high $\alpha$ and $\sigma$, while exhibiting low $\kappa$, but such a combination is inherently difficult to achieve.  In general, it has proven much more difficult to engineer high PF than to decrease $\kappa_{\text{lat}}$ even to the theoretical minimum given by the amorphous limit \cite{lowerlimit,nanostructure1,nanostructure3,lonepair1,lonepair2,resonantbonding,filledskut,snseorbitally}. The PF is intrinsically limited by the fact that $\sigma$ and $\alpha$ for a given material behave counteractively with respect to the carrier (doping) concentration. Conductivity can also be boosted by high carrier mobility, which occurs only in crystalline compounds. Unfortunately, high crystallinity often leads to high $\kappa$ as well, limiting the thermoelectric potential. Methods that suppress $\kappa_{\text{lat}}$, such as alloying, hierarchical nanostructuring, or lowering crystallinity often run the risk of damaging mobility, thereby decreasing the PF. A general consensus in the community is that systematic approaches for achieving high PF are of vital importance for the development of next-generation thermoelectrics.


One of the recent concepts for high PF focuses on the role of electronic bands that exhibit anisotropic features common in low-dimensional materials\cite{quantumwell,lowdimensional3d}, such as coexistence of flat and dispersive directions at the band edge. The idea behind this approach is that the flat portions provide high entropy due to high eDOS and leading to high Seebeck coefficient, while the dispersive portions facilitate high mobility due to light effective mass. Bulk materials with such features have been proposed theoretically, e.g. iron-based full-Heusler compounds of the Fe$_{2}$YZ type \cite{fe2yz}, but have not yet been proven useful.

It is in this context that we present our prediction of unprecedented $zT=5$ at 800 K in $n$-type full-Heusler Ba$_{2}$BiAu. This material is a member of the family of full-Heusler compounds recently discovered to be stable and to exhibit ultralow intrinsic lattice thermal conductivity arising from anharmonic rattling of heavy atoms \cite{ultralowheusler}. This compound also exhibits a coexistence of flat and dispersive directions at the valence band maximum (VBM), which suggests a possibility for a high $p$-type PF. Here, via parameter-free first-principles calculations based on the density-functional theory (DFT), we predict that Ba$_{2}$BiAu features high $n$-type PF arising from a {\it highly dispersive band\/} with multiple degenerate pockets at the conduction band minimum (CBM). By comparison, the $p$-type PF arising from the flat-and-dispersive, low-dimensional-type band at the VBM is significantly lower due to much increased phase space for electron scattering by acoustic phonons. We complement this observation with a case study of Fe$_{2}$TiSi, another full-Heusler compound that shares a qualitatively similar band structure, but with reversed locations of the highly dispersive and low-dimensional-type pockets. In both cases, we show that the highly dispersive band are more beneficial than the low-dimensional-type, suggesting that the more conventional concept of engineering multiply degenerate pockets with light carriers outperforms the benefits of highly anisotropic, flat-and-dispersive band structures.

\begin{figure}[tp]
\includegraphics[width=0.8 \linewidth]{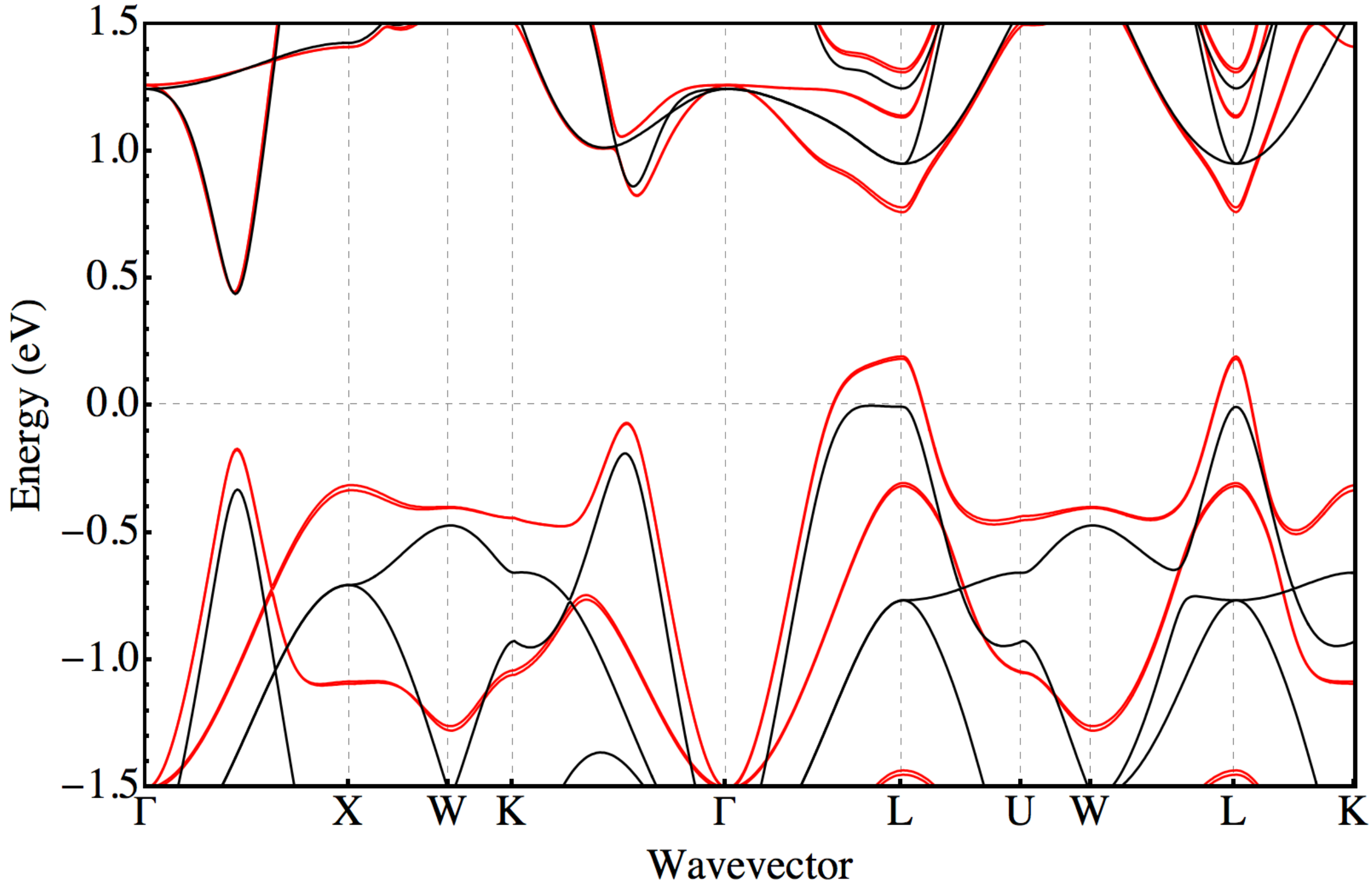}
\caption{(Color online) Electronic band structure of Ba$_{2}$BiAu with (red) and without (black) spin-orbit coupling, aligned at the CBM.} 
\label{fig:bandstructure}
\end{figure}

Fig. \ref{fig:bandstructure} shows the electronic band structure of Ba$_{2}$BiAu, as calculated using Quantum Espresso \cite{qespresso1,qespresso2} with norm-conserving pseudopotentials and the Perdew-Burke-Ernzerhof (PBE) exchange-correlation functional \cite{pbe,gga}; we confirmed that projector-augmented wave (PAW) \cite{paw} pseudopotentials lead to only minor changes in the band structure. The notable features of Fig.~\ref{fig:bandstructure} are a highly dispersive CBM pocket along the $\Gamma$-X direction and a flat-and-dispersive VBM pocket at the L-point. Both of these features are promising for high PF. First, the location of the CBM along the $\Gamma$-X direction gives it a sixfold pocket multiplicity, while the VBM pocket is fourfold degenerate. Pocket multiplicity, or band degeneracy, directly benefits the PF via the multi-band effect \cite{bandconvergence1} which increases conductivity without harming the Seebeck coefficient. No other portion of any band is within 0.3 eV of the CBM, making the pocket the sole contributor to electron transport up to 800 K. Second, the small effective mass of the CBM pocket limits the phase space for acoustic phonon scattering, which is often the most harmful carrier scattering process. We also note that spin-orbit coupling (SOC) induces a slight curving of the otherwise flat top-most valence band at the L-point and reduces the band gap, but it leaves the dispersive CBM pocket practically unchanged. Effective masses at the CBM are $m_{xx}=0.069$, $m_{yy}=m_{zz}=0.51$ without SOC and $m_{xx}=0.067$, $m_{yy}=m_{zz}=0.49$ with SOC. Since our main interest here is the performance of $n$-type material, and the SOC has negligible effect on both the CBM pocket and the phonon dispersion\cite{ultralowheusler}, we neglect SOC when calculating the electron-phonon scattering rates aside from correcting the band gap. In our transport calculations, we replace the non-SOC PBE gap of 0.44 eV with 0.56 eV obtained from the Tran-Blaha-modified Becke-Johnson potential (mBJ) \cite{mbj} with SOC.

We consider the two major scattering mechanisms at work in thermoelectric materials: electron-phonon scattering and ionized impurity scattering. For electron-phonon scattering, we explicitly calculate electron-phonon interaction matrix elements over a coarse $8\times8\times8$ $\textbf{k}$-point mesh using density-functional perturbation theory (DFPT) \cite{ponceprl,poncejcp}. We then use maximally localized Wannier functions \cite{mlwfcomposite,mlwfentangled,wannier90} for an efficient interpolation of the matrix elements onto a dense $40\times40\times40$ \textbf{k}-point mesh using the EPW software \cite{epw1,epw2,epw3}. This treatment accounts for both intravalley and intervalley scattering. Long-ranged polar optical scattering matrix elements are calculated separately and added to the total matrix elements on the dense mesh \cite{epwpolar}. The carrier lifetimes are obtained from the imaginary part of the electron self-energy, which is calculated by summing the matrix elements over the dense phonon mesh. For further theoretical details, refer to the supplementary information (SI) \cite{supplementary}. With lifetimes at hand, we employ the Boltzmann transport equation within the relaxation time approximation (RTA) to calculate $\sigma$ and $\alpha$:
	\begin{equation}\label{eq:sigma}
	\sigma=\frac{1}{\Omega N_{\mathbf{k}}} \sum_{\nu\mathbf{k}} (\tau v^{2})_{\nu\mathbf{k}}\left(-\frac{\partial f}{\partial \epsilon}\right)_{\epsilon=\epsilon_{\nu\mathbf{k}}},
	\end{equation}
	\begin{equation}\label{eq:alpha}
	\alpha=\frac{\sigma^{-1}}{\Omega TN_{\mathbf{k}}} \sum_{\nu\mathbf{k}} (\tau v^{2})_{\nu\mathbf{k}}(\epsilon_{\nu\mathbf{k}}-\epsilon_{\text{F}})\left(-\frac{\partial f}{\partial \epsilon}\right)_{\epsilon=\epsilon_{\nu\mathbf{k}}}.
	\end{equation}
Group velocity ($v$) and lifetime ($\tau$) depend on band $\nu$ and wavevector \textbf{k}. $\Omega$ is the primitive cell volume, $\epsilon_{\text{F}}$ is the Fermi level, and $f(\epsilon-\epsilon_{\text{F}})$ is the Fermi-Dirac distribution. The summations of Eqs. (\ref{eq:sigma}) and (\ref{eq:alpha}) are performed using BoltzTraP \cite{boltztrap} with pre-calculated electron-phonon lifetimes.  Recent work has shown that RTA with electron-phonon matrix elements calculated via DFPT and Wannier interpolation is capable of accurately reproducing experimental data on the electrical transport properties of good thermoelectric materials \cite{pbteabinitio,halfheuslersymmetry}. 

Mobility limited by ionized impurity scattering ($\mu_{\text{ii}}$) is estimated using the aMoBT software \cite{amobt}, which implements an improved form of the Brooks-Herring theory that takes into account screening and band non-parabolicity \cite{rode,ionizedimpurity}. The overall mobility ($\mu$) is obtained by combining $\mu_{\text{ii}}$ with $\mu_{\text{eph}}$ according to the Matthiesen's rule,
	\begin{equation}\label{eq:matthiesen}
	\mu^{-1}=\mu_{\text{eph}}^{-1}+\mu_{\text{ii}}^{-1},
	\end{equation}
and is used to rescale $\sigma$.

\begin{figure}[tp]
\includegraphics[width=1 \linewidth]{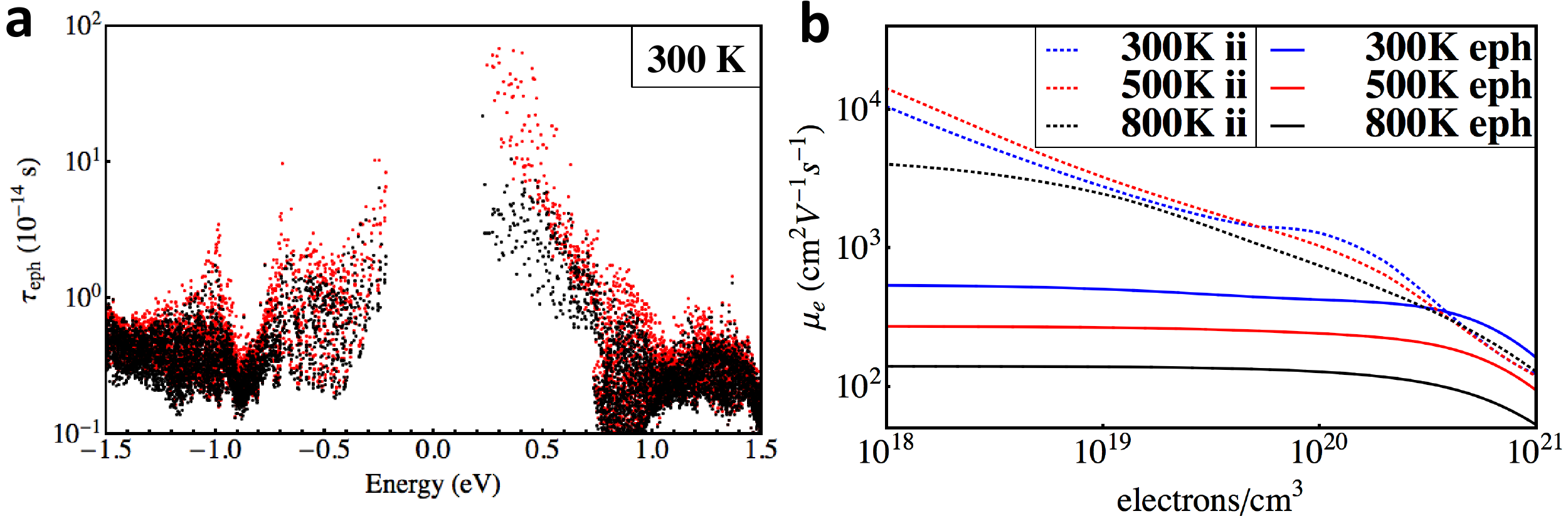}
\caption{(Color online) \textbf{(a)}  Energy-dependent electron-phonon scattering lifetimes at 300 K of Ba$_{2}$BiAu with (black) and without (red) polar optical scattering. \textbf{(b)} Electron mobility limited by electron-phonon scattering (eph, solid lines) and by ionized impurity scattering (ii, dashed lines) versus electron concentration.} 
\label{fig:eph}
\end{figure}

The calculated electron-phonon scattering lifetimes at 300 K are shown in Fig. \ref{fig:eph}(a). If only acoustic phonon scattering is included (red dots), electron lifetimes near the CBM are roughly two orders of magnitude higher than those of the hole states near the VBM. This shows that acoustic phonon scattering is indeed weak for the CBM states, as expected due to limited scattering phase space for a dispersive band. Inclusion of polar optical scattering reduces carrier lifetimes by an order of magnitude at the CBM (black dots). Polar optical scattering is then the dominant scattering mechanism for electrons at the CBM, as mobilities in Fig. \ref{fig:eph}(b) show. This mirrors the characteristics of other dispersive-band, high-mobility materials such as InSb and GaAs. Screening effects, which are not accounted for in our calculations and may become significant at higher carrier concentrations, would further increase the predicted carrier lifetimes and mobility. However, its effect will be counteracted by more intense electron-electron scattering. The contribution of ionized impurity scattering is for the most part small other than at very high doping concentrations. Hole mobility, as provided in the SI \cite{supplementary}, is significantly lower due to the flatter valence band.

\begin{figure}[tp]
\includegraphics[width=1 \linewidth]{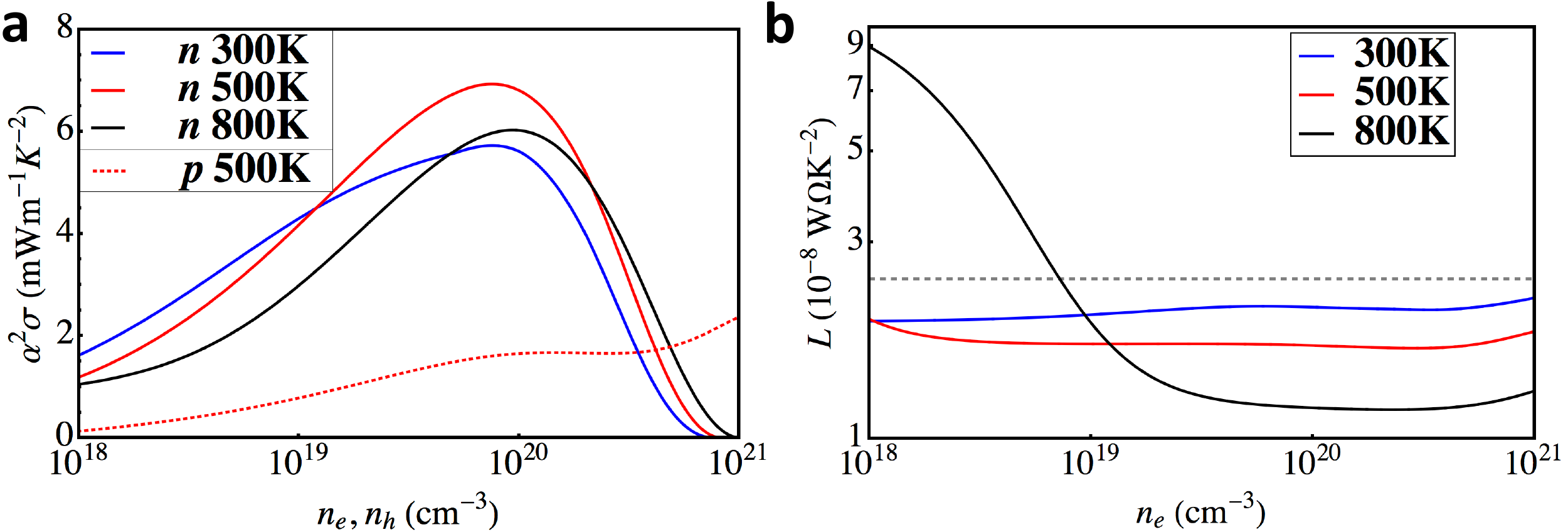}
\caption{(Color online) \textbf{(a)} The $n$-type PF and the 500 K $p$-type PF versus respective doping concentrations. \textbf{(b)} The $n$-type Lorenz number versus electron doping concentration, where the gray horizontal line indicates $L_{\text{WF}}$.}
\label{fig:power}
\end{figure}

The calculated PF tops out at 7 mW m$^{-1}$ K$^{-2}$ at 500 K (see Fig. \ref{fig:power}). This is among the highest $n$-type PFs for bulk semiconductors, approaching the current record value of 8 mW m$^{-1}$ K$^{-2}$ for the $p$-type half-Heusler compound NbFeSb at 500 K \cite{nbfesbpnas,halfheuslersymmetry}. The high PF of Ba$_{2}$BiAu is attributed in part to high conductivity arising from the sixfold pocket multiplicity and weak acoustic phonon scattering that elevates the electron lifetimes near the CBM. Also, the band generates a high $n$-type Seebeck coefficient which exceeds $-300$ $\mu$V K$^{-1}$ at 800 K (the plots of electrical conductivity and the Seebeck coefficient are shown in Fig. S5 of the SI \cite{supplementary}). The maximum PF requires degenerate doping into the conduction band, which is typical of phonon-limited electron transport. We find that ionized impurity scattering lowers the overall PF by 25\% at 300 K and by about 15\% at 500 K and 800 K.

Ultimately, high $zT$ requires high PF coupled with low thermal conductivity. Ba$_{2}$BiAu has already been predicted to possess ultralow lattice thermal conductivity \cite{ultralowheusler}: $\kappa_{\text{lat}}$ is 0.55 W m$^{-1}$ K$^{-1}$ at 300 K and 0.2 W m$^{-1}$ K$^{-1}$ at 800 K, close to the amorphous limit. Because of such low $\kappa_{\text{lat}}$, heat is mainly carried by electrons ($\kappa_{\text{e}}\ge1 $ W m$^{-1}$ K$^{-1}$). Since $\kappa_{\text{e}}>\kappa_{\text{lat}}$, small Lorenz number ($L$) is indispensable for the enhancement of $zT$. This statement can be generalized beyond Ba$_{2}$BiAu to any potential high $zT$ material since it must have low $\kappa_{\text{lat}}$ and high PF. As Fig. \ref{fig:power}(b) shows, our calculation indicates that throughout relevant doping ranges, $L$ is significantly lower than  the classical Wiedemann-Franz value $L_\text{WF}=\frac{\pi^{2}}{3}\frac{k_{\text{B}}^{2}}{e^{2}}$. Such a  deviation from the Wiedemann-Franz law is attributed to energy-dependent electron lifetime due to electron-phonon scattering. A simple illustration is provided by the non-degenerate approximation to the single parabolic band, according to which \cite{thermoelectrics}
	\begin{equation}\label{eq:lorenz}
	L=\frac{k_{\text{B}}^{2}}{e^{2}}\left(\frac{5}{2}+r\right)
	\end{equation}
where the energy-dependence of scattering time controls the constant $r$. For polar optical scattering, $r=\frac{1}{2}$, which leads to $L=\frac{3k_{\text{B}}^{2}}{e^{2}}< L_{\text{WF}}$. Under acoustic phonon scattering, $r=-\frac{1}{2}$ and $L$ would be even lower at $\frac{2k_{\text{B}}^{2}}{e^{2}}$. As the Fermi level moves towards the center of the gap with decreasing dopant concentration, $L$ increases above $L_{\text{WF}}$ as each electron carries more energy and bipolar transport takes effect [see Fig. \ref{fig:power}(b)]. Because electrons are much more mobile than holes, $L$ requires relatively little $n$-doping to sink below $L_{\text{WF}}$, to the benefit of $zT$.

\begin{figure}[tp]
\includegraphics[width=0.8 \linewidth]{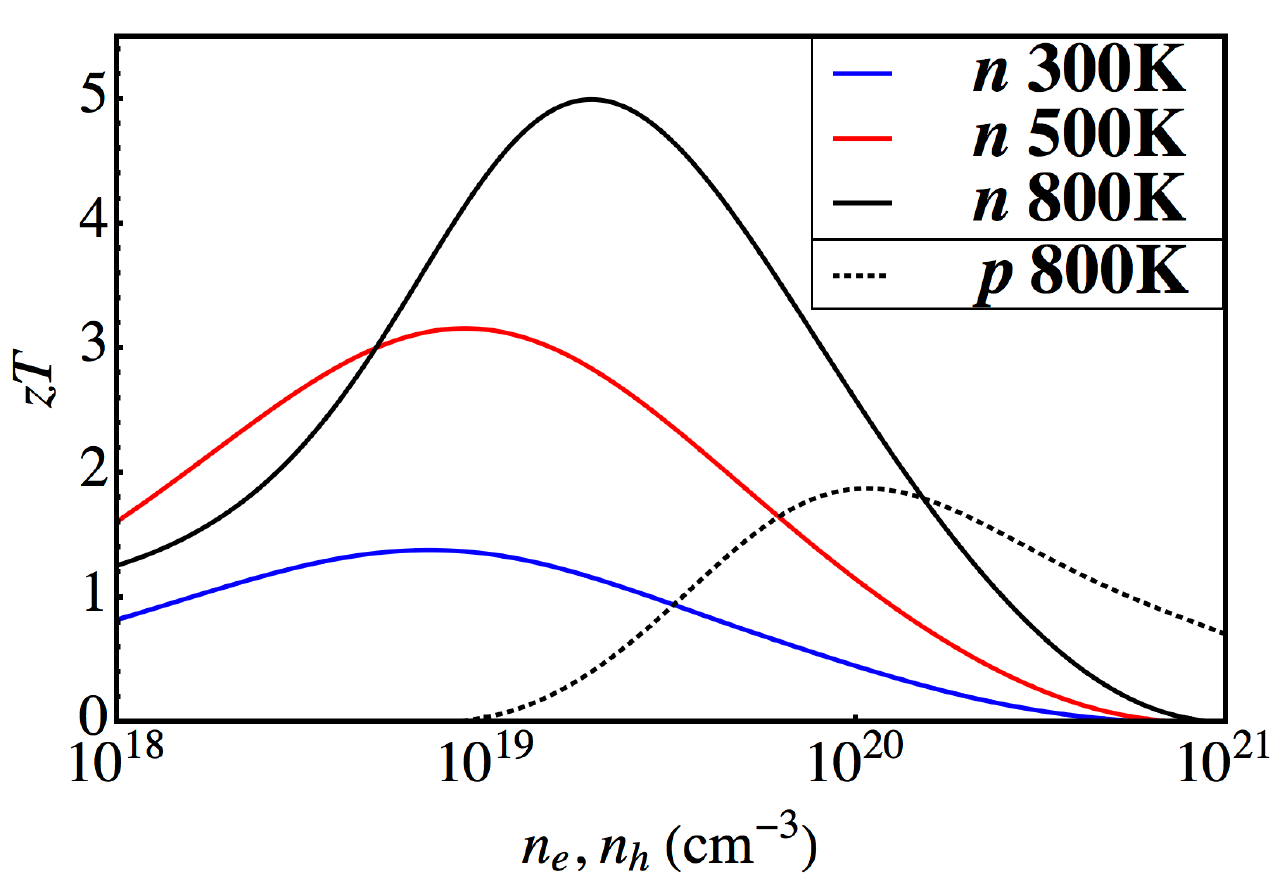}
\caption{(Color online) The $n$-type $zT$ the $p$-type $zT$ at 800 K versus respective doping concentrations.}
\label{fig:zt}
\end{figure}

The combination of high PF and low thermal conductivity in Ba$_{2}$BiAu leads to an unprecedentedly high $n$-type $zT=5$ at $800$ K. This value is higher than any other 3D bulk thermoelectric to date. At 300 K, we predict $zT=1.4,$ which would also be the highest for an $n$-type material at room temperature. Fig. \ref{fig:zt} shows that the peaks of $zT$ form at noticeably lower doping concentrations compared to those of the PF. Optimal $n_{e}$ for $zT$ in fact corresponds to the non-degenerate doping regime. The large difference between the optimal $n_{e}$ for the PF and $zT$ is reflective of the fact that $\kappa_{\text{e}}$ contributes heavily to the total thermal conductivity. If $\kappa_{\text{e}}>\kappa_{\text{lat}}$ and $L$ is relatively constant with $n_{e}$, $zT$ favors lower $n_{e}$ where the Seebeck coefficient is high ($-300$  $\mu$V K$^{-1}$ at 800 K). We note that our calculations result in $zT=0.8$ at 200 K and $zT=0.2$ at 100 K. Though not as remarkably high as the values we highlight, they are still respectable considering the very low temperatures.

We point out at this stage that Ba$_{2}$BiAu has previously been projected to have higher $p$-type PF than $n$-type PF due to the flat-and-dispersive valence band \cite{ultralowheusler}. According to our calculations, however, the highest $p$-type PF for Ba$_{2}$BiAu is merely 2.5 mW m$^{-1}$ K$^{-2}$ at 500 K and the highest $p$-type $zT$ is at best 2 at 800 K. We find that the flat portion of the band invites strong acoustic phonon scattering and this alone reduces hole lifetimes in the valence bands below the overall electron lifetimes in the conduction bands [see Fig. \ref{fig:eph}(a)]. Though SOC adds a small curvature to the VBM and therefore would somewhat improve hole lifetimes and the $p$-type properties, it is unlikely to make a significant difference. Additionally, the valence band only has fourfold pocket multiplicity at the $L$-point. Shorter hole lifetimes, fewer pockets, and lower group velocities force one to rely on very high hole doping ($n_{h}>10^{21}$ cm$^{-3}$) to achieve sufficiently high values of $\sigma$ in $p$-type samples. However, this reduces the Seebeck coefficient to approximately 80~$\mu$V~K$^{-1}$ at 500 K, resulting in poor PF. As a side-remark, such a high hole doping requirement may also make synthesis difficult. These observations suggest that highly dispersive bands that minimize acoustic electron-phonon scattering can easily outperform flat-and-dispersive bands.

\begin{figure}[tp]
\includegraphics[width=0.8 \linewidth]{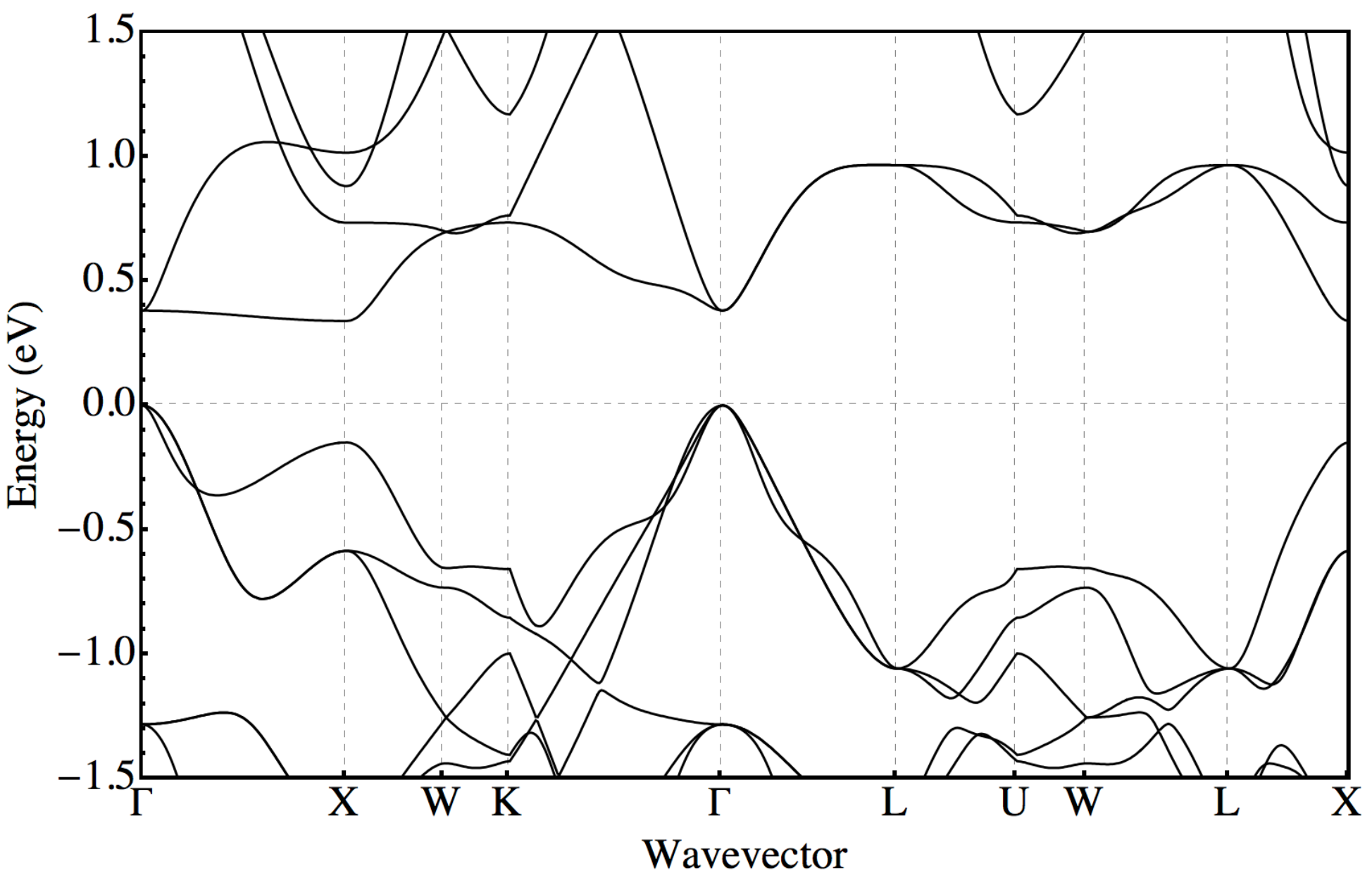}
\caption{The band structure of Fe$_{2}$TiSi.} 
\label{fig:fe2tisiband}
\end{figure}

To support this observation and to further generalize it, we consider Fe$_{2}$TiSi, another full-Heusler compound which has been previously suggested as a possible high-performance themoelectric \cite{fe2yz,fe2tisintype}. As Fig. \ref{fig:fe2tisiband} shows, its conduction bands represent a quintessential example of a flat-and-dispersive band structure. Accordingly, the $n$-type PF has drawn popular interest, though previous theoretical studies have resorted to various approximations for computing carrier scattering rates \cite{fe2yz,fe2tisintype}. Using the fully first-principles approach for electron-phonon scattering adopted in this work, we find that the $n$-type thermoelectric performance is in fact rather modest and comparable to the $p$-type PF arising from the dispersive-only valence bands. Detailed discussion of the results of our calculations for Fe$_{2}$TiSi is given in the SI (see Figs. S7$\sim$S10) \cite{supplementary}. We find that the electron lifetimes are much lower than the hole lifetimes due to strong acoustic phonon scattering accommodated by the flat conduction band. Also, the flat-and-dispersive conduction bands do not offer an intrinsic advantage in the Seebeck coefficient relative to the purely dispersive valence bands. The predicted maximum $n$-type PF arising from the flat-and-dispersive bands in fact relies on high electrical conductivity promoted by the high number of dispersive pockets at the CBM (three at X and additional doubly degenerate band at $\Gamma$). In contrast, the VBM is only triply degenerate at $\Gamma$ but still has a comparable PF, and would easily outperform the $n$-type PF if it had a higher degeneracy factor (as occurs for the dispersive $n$-type band in Ba$_{2}$BiAu).

In summary, we predict that $n$-type Ba$_{2}$BiAu has the potential to achieve $zT=5$ at 800 K and $1.4$ at 300 K due to a rare complementation of very high PF, ultralow lattice thermal conductivity, and a small Lorenz number. The predictions of this work are offered for experimental synthesis and characterization, which would constitute a proof of principle that the long-sought $zT$ above 4 is attainable. We trace the high $n$-type PF of Ba$_{2}$BiAu to a highly dispersive CBM pocket along the $\Gamma$-X direction. In contrast, the flat-and dispersive valence band of Ba$_{2}$BiAu allows much smaller $p$-type PF values due to strong acoustic phonon scattering. A similar trend is predicted in Fe$_{2}$TiSi, where the flat-and-dispersive band is at the CBM. The commonalities between Ba$_{2}$BiAu and Fe$_{2}$TiSi strongly suggest that highly dispersive bands with multiple pockets ought to be the main focus in the quest for high thermoelectric performance in bulk materials, while flat-and-dispersive, low-dimensional band structures usually invite strong electron-phonon scattering and reduced PFs. From this perspective, it is only fitting that the highest known PF for a semiconductor occurs in NbFeSb, which has a very weak deformation potential \cite{halfheuslersymmetry} but also eight dispersive pockets, only its high $\kappa_{\text{lat}}$ limits its overall thermoelectric efficiency. We hope that these examples will stimulate further effort in search of highly dispersive carrier pockets occurring in low-$\kappa_{\text{lat}}$ materials. 


The authors acknowledge financial support from the National Science Foundation Grant DMR-1611507. This research used resources of the National Energy Research Scientific Computing Center, a DOE Office of Science User Facility supported by the Office of Science of the U.S. Department of Energy under Contract No. DE-AC02-05CH11231. The authors would also like to thank Dr. Alireza Faghaninia of Lawrence Berkeley National Laboratory for helpful discussions on aMoBT.

\bibliography{references}

\end{document}


\title{Supplementary Materials: High $n$-type thermoelectric power factor and efficiency in Ba$_{2}$BiAu from a highly dispersive band}
	\author{Junsoo Park}
	\affiliation{Department of Materials Science \& Engineering, UCLA, Los Angeles, CA 90095, USA} 
	\affiliation{Department of Applied Physics, Yale University, New Haven, CT 06511, USA} 
	\affiliation{Energy Sciences Institute, Yale University, West Haven, CT 06516, USA} 	
	\author{Yi Xia}
	\affiliation{Nanoscience and Technology Division, Argonne National Laboratory, Argonne, IL  60439, USA}
	\author{Vidvuds Ozoli\c{n}\v{s}}
	\affiliation{Department of Applied Physics, Yale University, New Haven, CT 06511, USA} 	
	\affiliation{Energy Sciences Institute, Yale University, West Haven, CT 06516, USA} 	
	\date{\today} 
	\pacs{******}
	\maketitle

\section{Crystal \& Electronic Structures}

Ba$_{2}$BiAu is full-Heusler compound, which can be described as a rock-salt structure of Bi and Au with all eight of its diamond sublattice positions occupied by Ba. Its primitive cell is that of face-centered cubic with additional atoms in the interior. Relaxation of crystal structure and self-consistent calculation of electronic structures of the primitive cell are performed using two different schemes: using norm-conserving (NC) pseudopotentials on Quantum Espresso (QE), hereafter NC-QE, and using PAW pseudopotentials on Vienna \textit{Ab Initio} Simulations Package (VASP) \cite{vasp1,vasp2,vasp3,vasp4}, hereafter PAW-VASP. PBE exchange-correlation functional is used for both cases. Plane-wave energy cut-off of 120 Ry and energy convergence threshold of $10^{-8}$ Ry are used for all NC-QE calculations. Plane-wave energy cut-off of 600 eV and energy convergence threshold of $10^{-7}$ eV are used for all PAW-VASP calculations. Convergence with respect to \textbf{k}-point mesh is safely ensured.

\begin{figure}[bp]
\includegraphics[width=0.7 \linewidth]{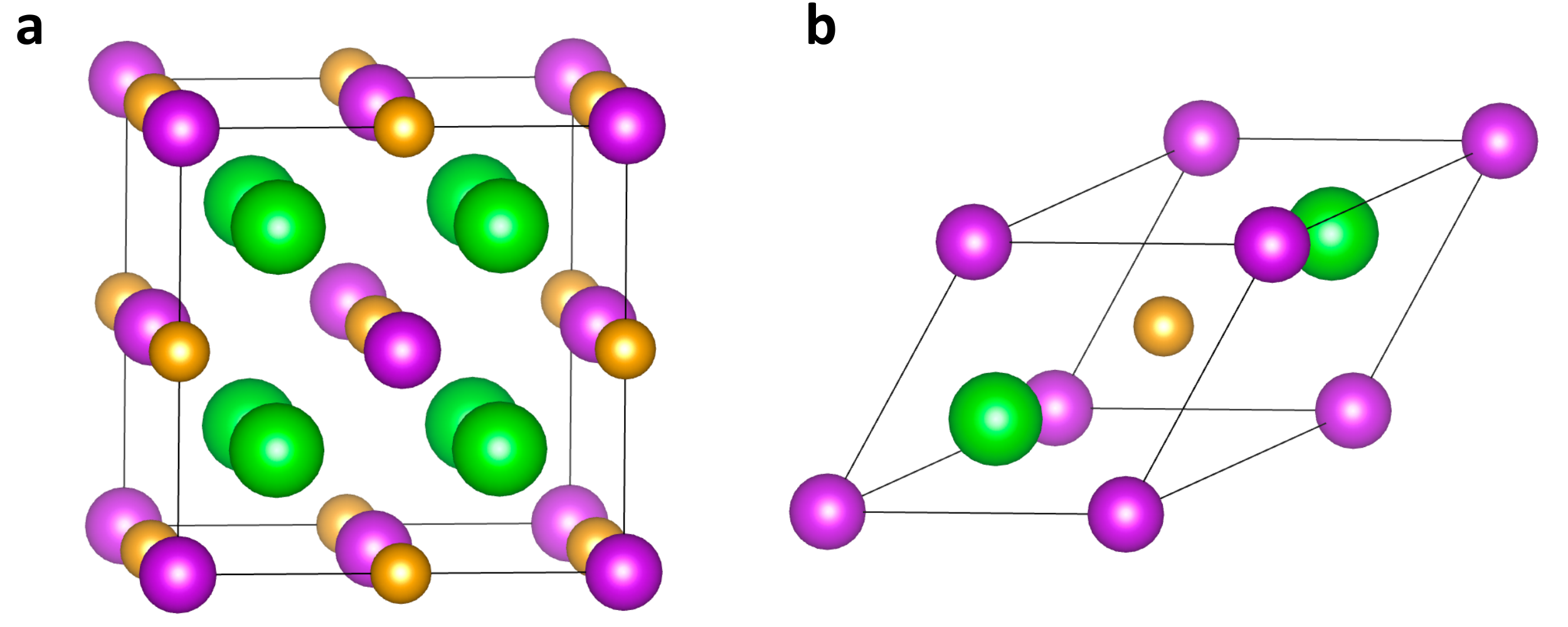}
\caption{(Color online) \textbf{a)} The unit cell. \textbf{b)} The primitive cell. Ba is green, Bi is purple, and Au is gold.} 
\label{fig:unitcell}
\end{figure}

The ground-state lattice parameter of Ba$_{2}$BiAu as calculated by NC-QE is 8.30 \r{A}, and that by PAW-VASP is 8.42 \r{A}. Ba$_{2}$BiAu has not been synthesized at all to date, so experimental data are unavailable for comparison. NC-QE and PAW-VASP yield more or less the same band structures for both compounds near their energy gaps. The 0 K band gap of Ba$_{2}$BaAu is 0.44 eV when calculated with plain NC-QE and 0.45 eV when calculated with PAW-VASP. With Heyd-Scuzeria-Ernzerhof (HSE) correction \cite{hse1,hse2}, the gap increases to 0.74 eV. With spin-orbital coupling but without HSE06, $\Delta E_{g}$ is 0.25 eV. The most accurate value is likely predicted by the mBJ functional with SOC, which yields $\Delta E_{g}$ of 0.56 eV. It is this value we use for transport calculations.

Ba$_{2}$BiAu features pipe-like isoenergy surface at the valence band maximum, reflective of its flat-and-dispersive nature. Effective mass ($m$) is large along the pipes, corresponding to the flat direction, and small around the pipes. On the other hand, the isoenergy surface at the CBM is a set of oblates poised in the middle of the $\Gamma$-X directions. These oblates reflect that $m$ is very small longitudinally and larger but still small transversely. Overall, the oblates are characteristics of dispersive pockets.

\begin{figure}[tp]
\includegraphics[width=0.7 \linewidth]{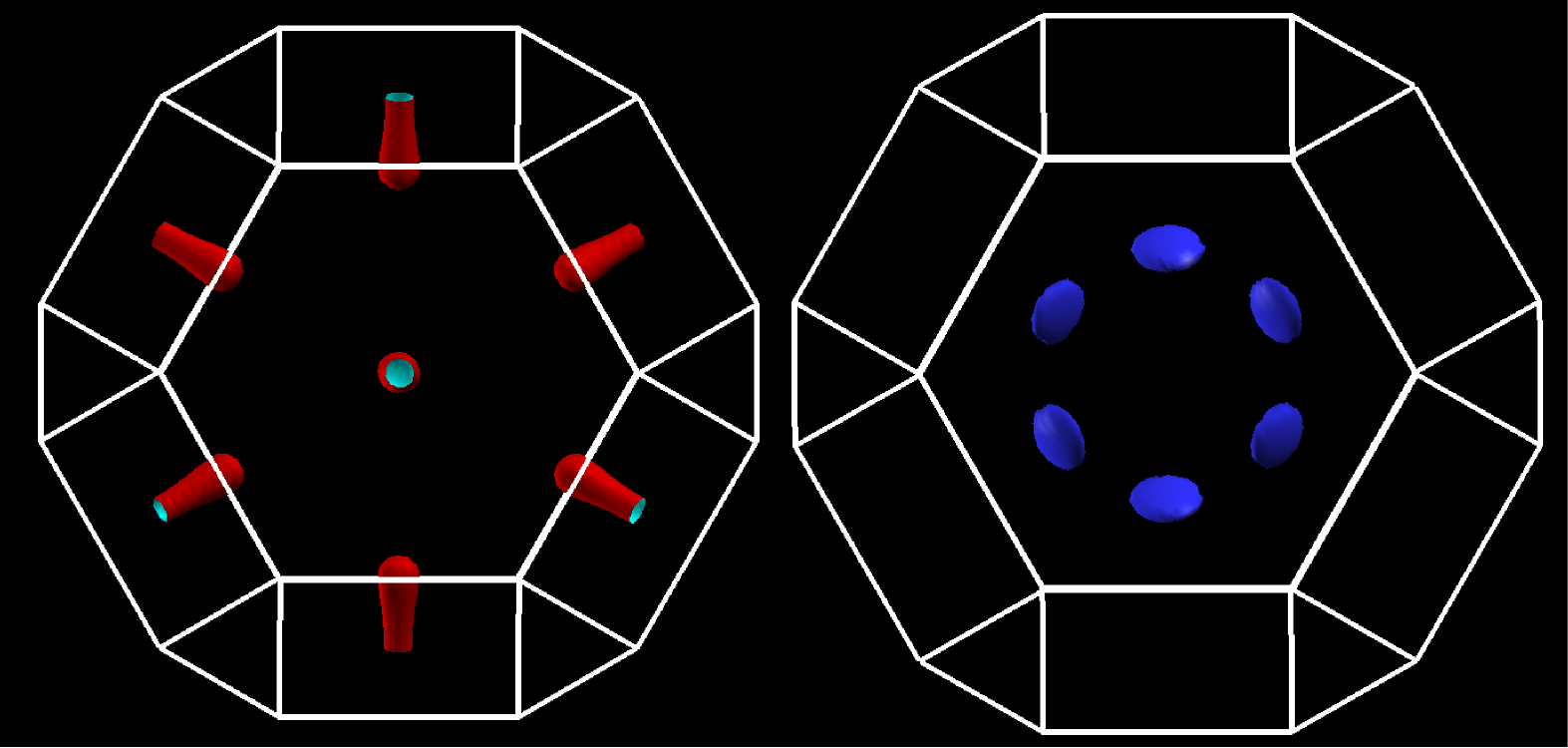}
\caption{(Color online) Isoenergy surfaces of the topmost valence band (left) and lowermost conductions band (right) of Ba$_{2}$BiAu near their extrema.}
\label{fig:ba2biaufermisurface}
\end{figure}

\begin{figure}[bp]
\includegraphics[width=0.9 \linewidth]{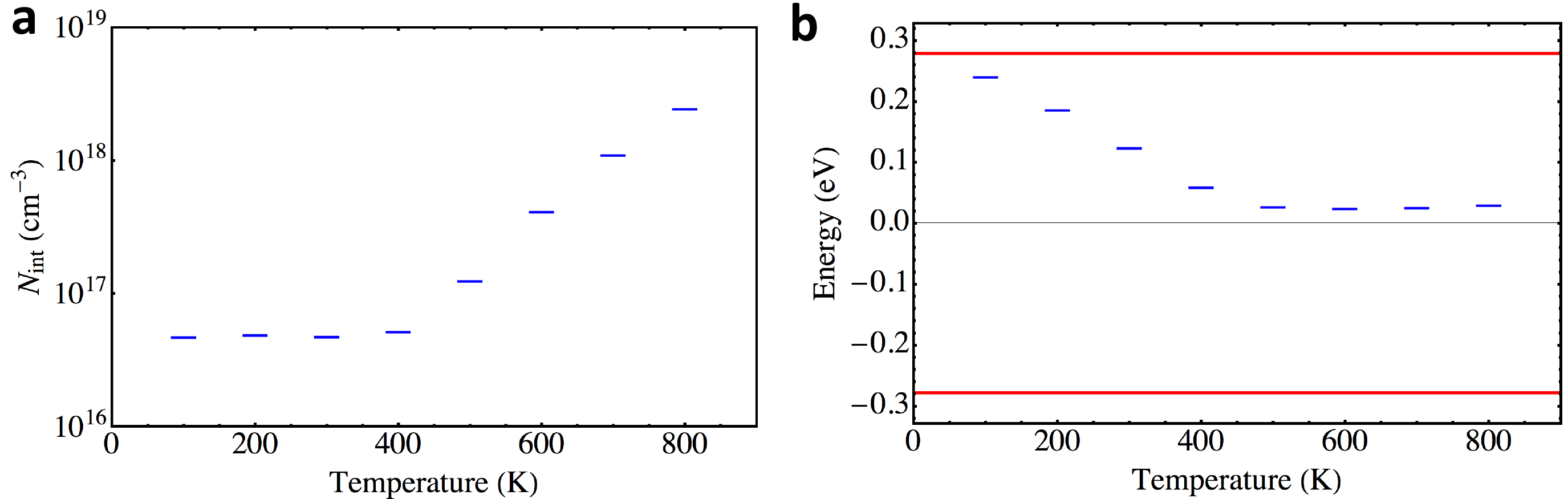}
\caption{(Color online) \textbf{a)} Intrinsic carrier concentration. \textbf{b)} Intrinsic Fermi level, where the horizontal red lines indicate the VBM (lower) and CBM (upper).}
\label{fig:intrinsic}
\end{figure}

Intrinsic carrier concentration ($N_{\text{int}}$) and intrinsic Fermi levels ($E_{\text{int}}$) are shown in Fig. \ref{fig:intrinsic}. At low temperatures, due to the much more dispersive conduction band, $E_{\text{int}}$ sits close to the conduction band. As the temperature rises, $E_{\text{int}}$ approaches the center of the gap.

\section{Electron-phonon scattering}

Since an accurate description of electron-phonon scattering is critical for thermoelectric performance, we perform rigorous matrix element calculations. Electron-phonon interaction matrix elements and corresponding electron self-energies are treated within the framework of harmonic phonon perturbation to electronic states. Electron-phonon scattering rate (inverse lifetime) dependent on band $\nu$ and phonon mode $\lambda$ is calculated as
\begin{equation}\label{eq:ephtau}
\tau_{\text{eph},\lambda\nu\mathbf{k}}^{-1}=\frac{2}{\hbar}\Sigma_{\text{Im},\nu\lambda\mathbf{k}}
\end{equation}
Here, $\Sigma_{\text{Im}}$ is the imaginary part of electron self-energy, a.k.a., electron linewidth. Within density functional perturbation theory, it is defined as
\begin{widetext}
\begin{equation}\label{eq:imaginarypart}
\begin{aligned}
	\Sigma_{\text{Im},\nu\lambda\mathbf{k}}=\frac{\pi}{N_{\mathbf{q}}}\sum_{\nu'\mathbf{q}} \left| g_{\nu'\nu\lambda}(\mathbf{k,q})\right|^{2}\times & [(b(\omega_{\lambda\mathbf{q}},T)+f(\epsilon_{\nu'\mathbf{k}+\mathbf{q}},\epsilon_{\text{F}},T))\delta(\epsilon_{\nu\mathbf{k}}+\omega_{\lambda\mathbf{q}}-\epsilon_{\nu'\mathbf{k}+\mathbf{q}})d\epsilon \\
	&+(b(\omega_{\lambda\mathbf{q}},T)+1-f(\epsilon_{\nu'\mathbf{k}+\mathbf{q}},\epsilon_{\text{F}},T)) \delta(\epsilon_{\nu\mathbf{k}}-\omega_{\lambda\mathbf{q}}-\epsilon_{\nu'\mathbf{k}+\mathbf{q}})d\epsilon] \\
\end{aligned}
\end{equation}
\end{widetext}
where $\omega$ is phonon frequency, $N_{\mathbf{q}}$ is the number of $\textbf{q}$-points, $\delta$ are energy-conserving delta functions, $b$ is phonon population given by the Bose-Einstein distribution, and $f$ is electron population given by the Fermi-Dirac distribution. The most important ingredient in the evaluation of $\Sigma_{\text{Im}}$ is mode-and-band-resolved electron-phonon interaction matrix elements:
\begin{equation}\label{eq:ephmatrix}
g_{\nu'\nu\lambda}(\mathbf{k,q})=\sqrt{\frac{1}{2m\omega_{\lambda\mathbf{q}}}}\left\langle \psi_{\nu'\mathbf{k}+\mathbf{q}}^{(0)} \right| \partial V_{\lambda\mathbf{q}} \left| \psi_{\nu\mathbf{k}}^{(0)}\right\rangle.
\end{equation}

\begin{figure}[bp]
\includegraphics[width=1 \linewidth]{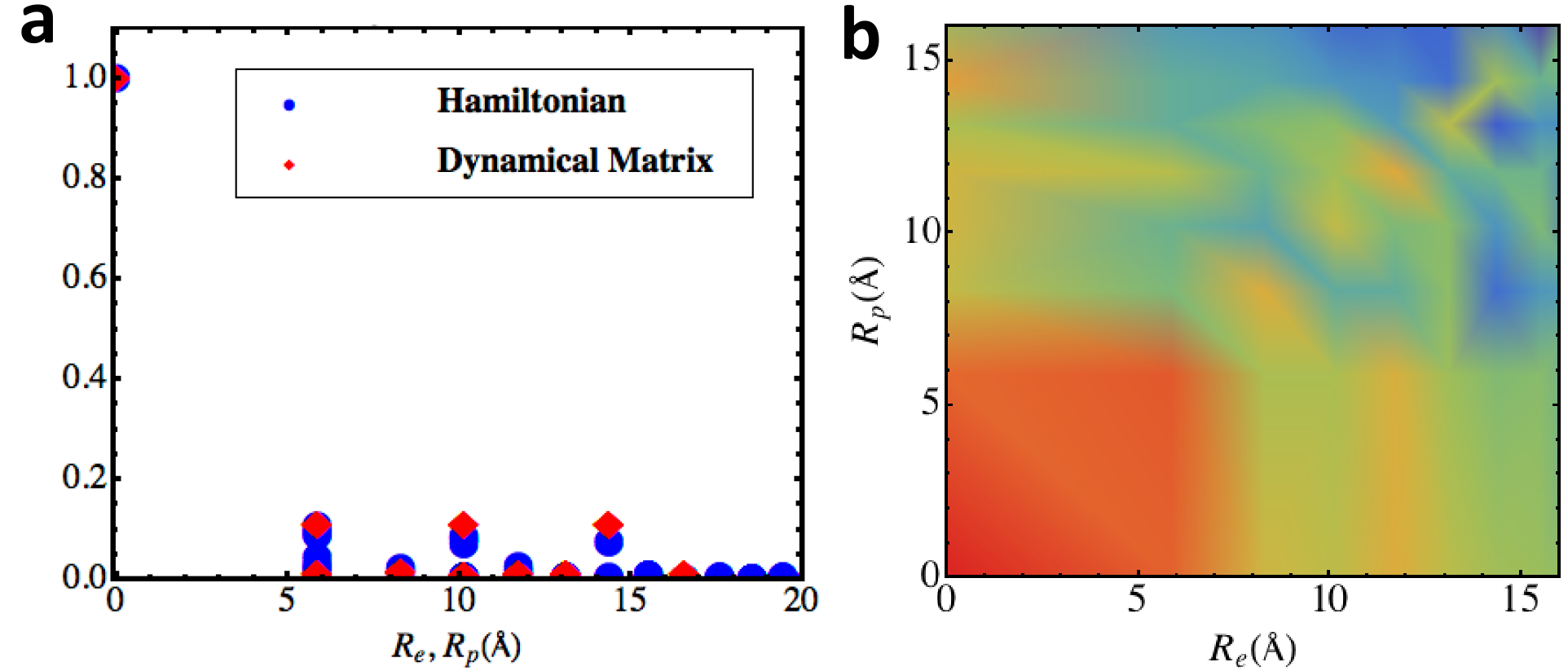}
\caption{(Color online) \textbf{a)} Localization of Wannier Hamiltonian and dynamical matrix elements, normalized to 1, of Ba$_{2}$BiAu. \textbf{b)} Log-scale decay of electron-phonon interaction matrix elements.}
\label{fig:decay}
\end{figure}

In the above expression, $\psi^{(0)}$ is the ground-state wavefunction and $\partial V$ is the perturbation due to atomic displacement by harmonic phonons. Since explicit calculation of a matrix element of the form of Eq. \ref{eq:ephmatrix} is computationally expensive, a reasonable approach is to compute it on a coarse mesh and then to interpolate it onto a much denser mesh throughout the Brillouin zone. To accomplish this, we use the EPW software, which performs the interpolation via maximally localized Wannier functions (MLWF). Polar optical scattering cannot be treated with Wannier interpolation due to the long-ranged nature of the interaction and the localized nature of MLWF. Therefore, they are calculated separately and then added to the total matrix elements on the dense mesh. The long-ranged polar optical scattering matrix elements are given by \cite{epwpolar}
\begin{widetext}
\begin{equation}\label{eq:polarmatrix}
g_{\nu'\nu\lambda}^{L}(\mathbf{k,q})=i\sum_{a}\sqrt{\frac{1}{2M_{a}\omega_{\lambda\mathbf{q}}}}\sum_{\mathbf{G}\ne-\mathbf{q}} \frac{(\mathbf{q}+\mathbf{G})\cdot Z^{*}\cdot\mathbf{e}_{a\mathbf{\lambda\mathbf{q}}}}{(\mathbf{q}+\mathbf{G})\cdot\epsilon^{\infty}\cdot(\mathbf{q}+\mathbf{G})} \left\langle \psi_{\nu'\mathbf{k}+\mathbf{q}}^{(0)} \right| e^{i(\mathbf{q}+\mathbf{G})\cdot\mathbf{r}} \left| \psi_{\nu\mathbf{k}}^{(0)}\right\rangle,
\end{equation}
\end{widetext}
where $Z^{*}$ is the Born effective charge tensor, $\epsilon^{\infty}$ is the high-frequency dielectric permittivity tensor, $M_{a}$ is the mass of atom $a$, \textbf{e} is phonon eigenvector, \textbf{G} is the reciprocal lattice vector, and \textbf{r} is atomic position vector.

\begin{figure}[tp]
\includegraphics[width=0.7 \linewidth]{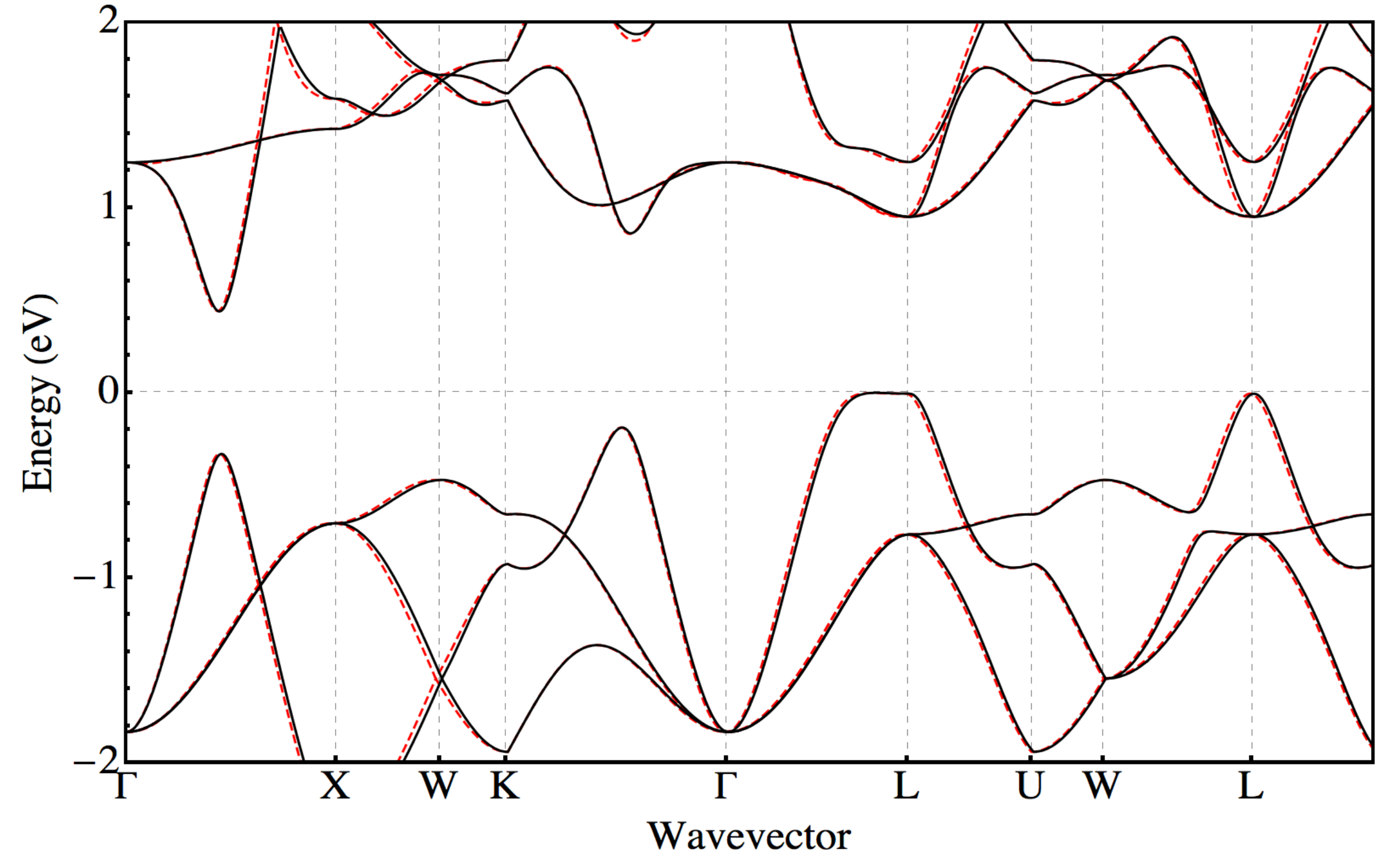}
\caption{(Color online) Wannier-interpolated band structure (red, dashed) overlaid on self-consistent band structure (black).}
\label{fig:wannierband}
\end{figure}

Since EPW runs in conjunction with QE and is compatible only with NC pseudopotentials, NC-QE rather than VASP is our DFT tool of choice for all calculations involved in electron-phonon scattering. In preparation for electron-phonon calculations, we first perform structural relaxation and self-consistent energy minimization. Convergence is safely achieved with a $40\times40\times40$ \textbf{k}-point mesh. We then compute harmonic phonons and their interaction matrices with electrons using a $8\times8\times8$ \textbf{k}-point mesh for electrons and a $4\times4\times4$ \textbf{q}-point mesh for phonons by DFPT. Finally, we interpolate the matrix elements onto $40\times40\times40$ \textbf{k}-point mesh and \textbf{q}-point mesh. We confirm the localization of the Wannier Hamiltonian, the real-space dynamical matrix and electron-phonon interaction matrix elements from their decay over short ranges, as shown in Fig. \ref{fig:decay}. Proper localization indicates accurate interpolation. The quality of Wannier interpolation is confirmed by the excellent matching of the interpolated band structure and the self-consistently calculated band structure (see Fig. \ref{fig:wannierband}). Using DFPT, we also obtain Born effective charges of $2.77$, $-4.45$, and $1.14$ for Ba, Bi and Au, respectively. They add up to 0 but for a very small residual. Deviations from the nominal charges may be caused potentially by hybridization of orbitals. The static dielectric permittivity is 21.24.

\begin{figure}[tp]
\includegraphics[width=0.9 \linewidth]{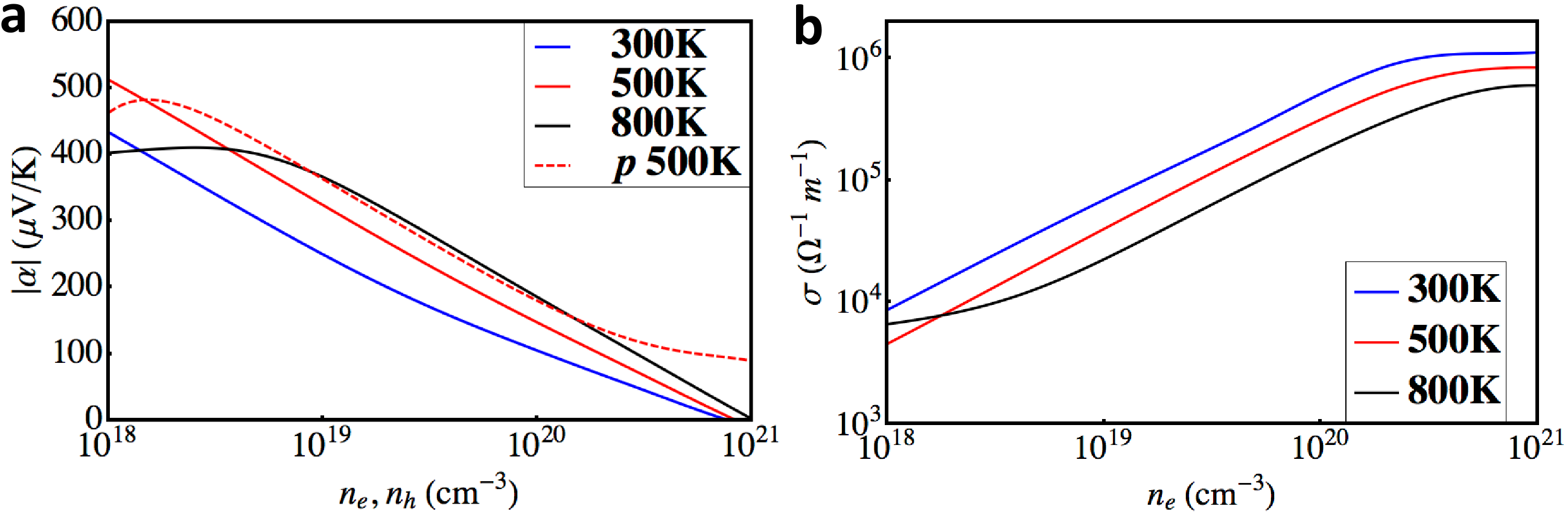}
\caption{(Color online) \textbf{a)} The absolute $n$-type Seebeck coefficients and $p$-type Seebeck coefficient at 500 K versus the respective doping concentrations. \textbf{b)} $n$-type conductivity and $p$-type conductivity at 500 K versus the respective doping concentrations. \textbf{c)} The Lorenz number in the hole doping regime. \textbf{d)} Hole mobility versus hole concentration.}
\label{fig:transport}
\end{figure}

Carrier conductivity and the Seebeck coefficients are provided in Fig. \ref{fig:transport} to complement the discussion of PF in the main text. Note that the Seebeck coefficient of $n$-type is similar to if not higher than that of $p$-type at respectively optimal carrier concentrations. This indicates that purely dispersive bands are capable of generating competitive Seebeck coefficient. Hole mobility and the Lorenz number in the hole-doping regime are also provided for reference. It is evident that neither is as favorable for $zT$ as their $n$-type counterparts for electrons due to the flatter valence band.

\section{Lattice Thermal Conductivity}

\begin{figure}[tp]
\includegraphics[width=0.6 \linewidth]{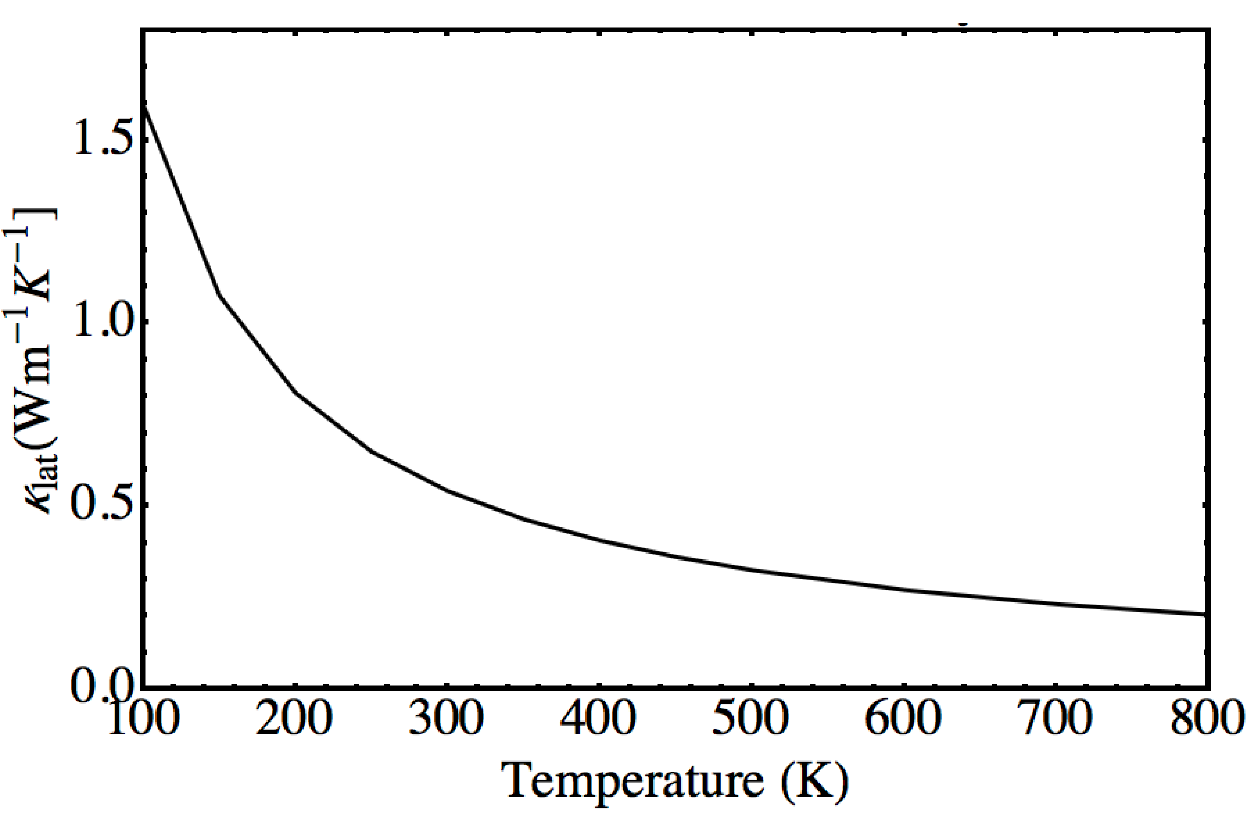}
\caption{Lattice thermal conductivity of Ba$_{2}$BiAu.} 
\label{fig:klat}
\end{figure}

Phonons and intrinsically ultralow lattice thermal conductivity of Ba$_{2}$BiAu have been previously investicated using state-of-the-art computational methods \cite{ultralowheusler}. The compressive sensing lattice dynamics technique \cite{csld} was used to calculate anharmonic interatomic force constants, and ShengBTE \cite{shengbte} was used for iterative Boltzmann transport calculations. We follow the same approach and arrive at identical results. For convenience, we provide our reproduction of the $\kappa_{\text{lat}}$ prediction in Fig. \ref{fig:klat}.

\section{Case study: F\lowercase{e}$_{2}$T\lowercase{i}S\lowercase{i}}

We look to Fe$_{2}$TiSi for supporting results and discussions because it features both a flat-and-dispersive band structure and a dispersive-only band structure, as shown in the main text. The conduction bands have a very pronounced flat portion along $\Gamma$-X and five dispersive pockets, one isotropic and four anisotropic. Meanwhile, the three valence bands are at $\Gamma$. The compound serves as a good platform for a comparing the performance of the two distinctly different band structures. We follow the same set of methods as appears in this paper to calculate its band structure, electron-phonon scattering, and transport properties.

\begin{figure}[bp]
\includegraphics[width=0.7 \linewidth]{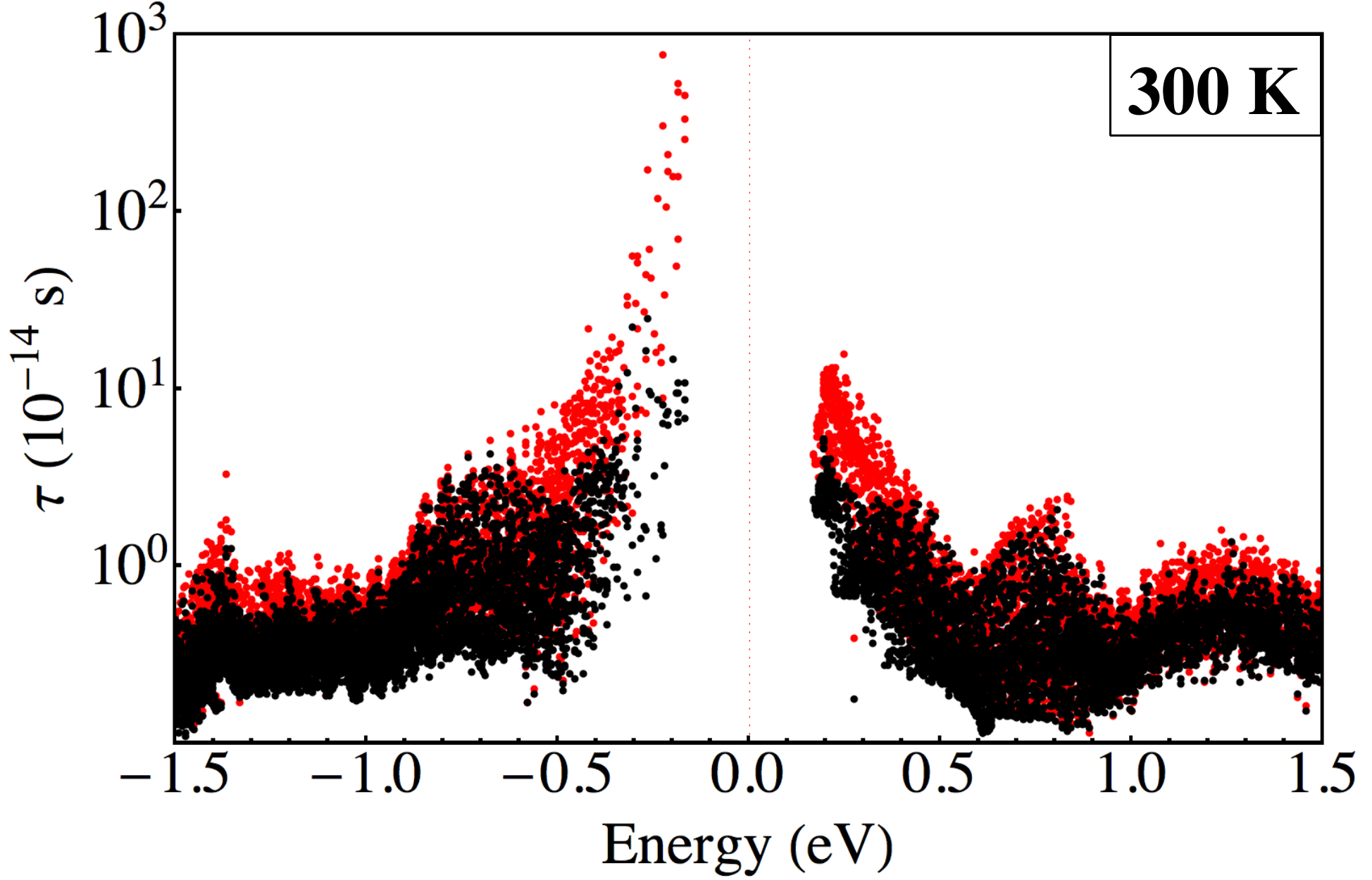}
\caption{Carrier lifetimes of Fe$_{2}$TiSi with (black) and without (red) polar optical scattering.} 
\label{fig:fe2tisitau}
\end{figure}

\begin{figure}[tp]
\includegraphics[width=0.7 \linewidth]{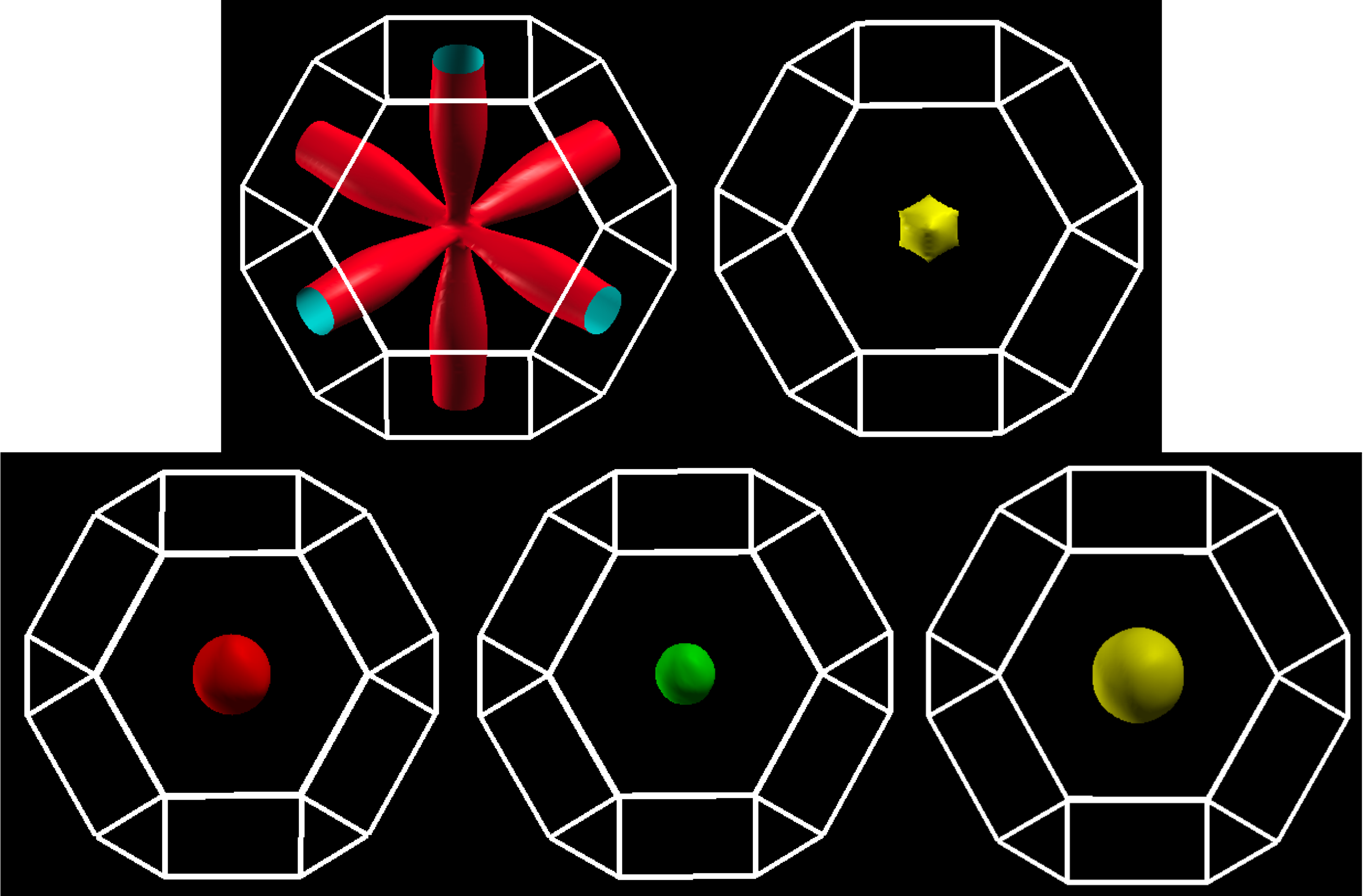}
\caption{The isoenergy surfaces at the CBM (top two) and the VBM (bottom three).} 
\label{fig:fe2tisifermi}
\end{figure}

Calculated lifetimes at 300 K are plotted in Fig. \ref{fig:fe2tisitau}. Without polar optical scattering, the VBM of Fe$_{2}$TiSi has even longer (hole) lifetimes than the CBM of Ba$_{2}$BiAu. This likely due to even smaller phase space offered by only the three valence bands as opposed to the six conduction band pockets in Ba$_{2}$BiAu. Because of this, polar optical scattering makes an even larger contribution in Fe$_{2}$TiSi than in Ba$_{2}$BiAu, reducing the lifetimes at the VBM by more than an order of magnitude. Meanwhile, the CBM consistently features short lifetimes, very similar in scale to those of the VBM of Ba$_{2}$BiAu without SOC. Again, a very large phase space for acoustic phonon scattering is at work. These phenomena can be understood from the isoenergy surfaces at the two band edges (see Fig. \ref{fig:fe2tisifermi}). The CBM features one surface resembling very pronounced orthogonal pipes, indicative of the anisotropically flat-and-dispersive nature, and a spheroidal surface corresponding to extra dispersive pockets at $\Gamma$. In contrast, all three valence bands are isotropically parabolic, as the three spherical surfaces reflect. These indicate arge and small acoustic phonon scattering phase spaces for the CBM and the VBM, respectively.

\begin{figure}[tp]
\includegraphics[width=0.8 \linewidth]{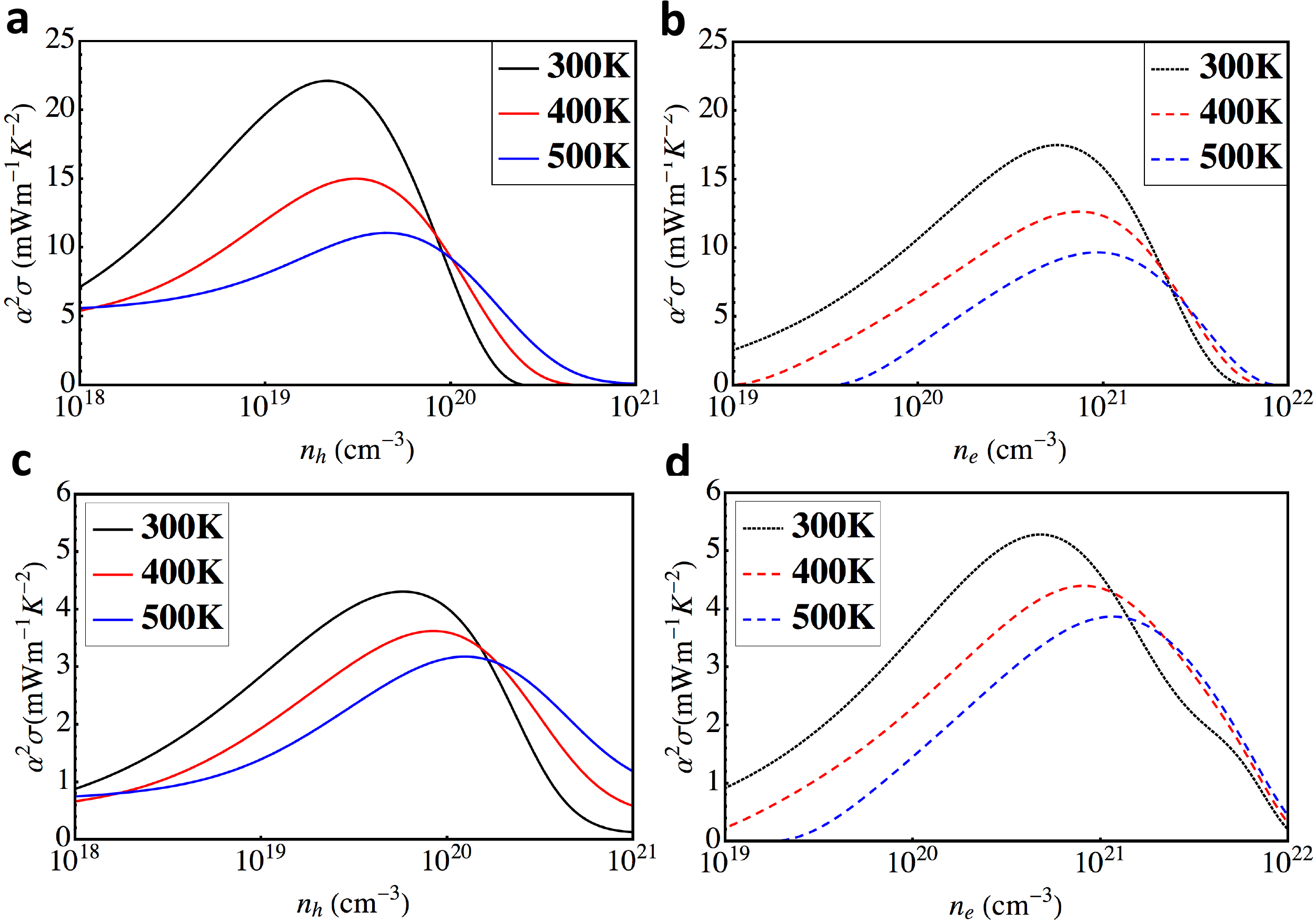}
\caption{The thermoelectric power factor without polar optical scattering of \textbf{a)} $p$-type and \textbf{b)} $n$-type Fe$_{2}$TiSi, and that with polar optical scattering of \textbf{c)} $p$-type and \textbf{d)} $n$-type Fe$_{2}$TiSi.} 
\label{fig:fe2tisipower}
\end{figure}

The resulting thermoelectric PF are shown in Fig. \ref{fig:fe2tisipower}. It is evident that the $p$-type PF is somewhat higher without polar optical scattering due to very long hole lifetimes, while the $n$-type PF is somewhat higher with polar optical scattering due to subdued hole lifetimes. Because polar optical scattering is such a dominant process, much more so than in Ba$_{2}$BiAu, any screening provided by free carriers would considerably improve polar-optical-limited hole lifetimes and the $p$-type PF. Such an improvement can be expected since carrier concentrations are high for thermoelectrics. Overall, the $p$-type performance delivered by the dispersive-only valence bands is at least comparable to the $n$-type performance. 

\begin{figure}[tp]
\includegraphics[width=0.8 \linewidth]{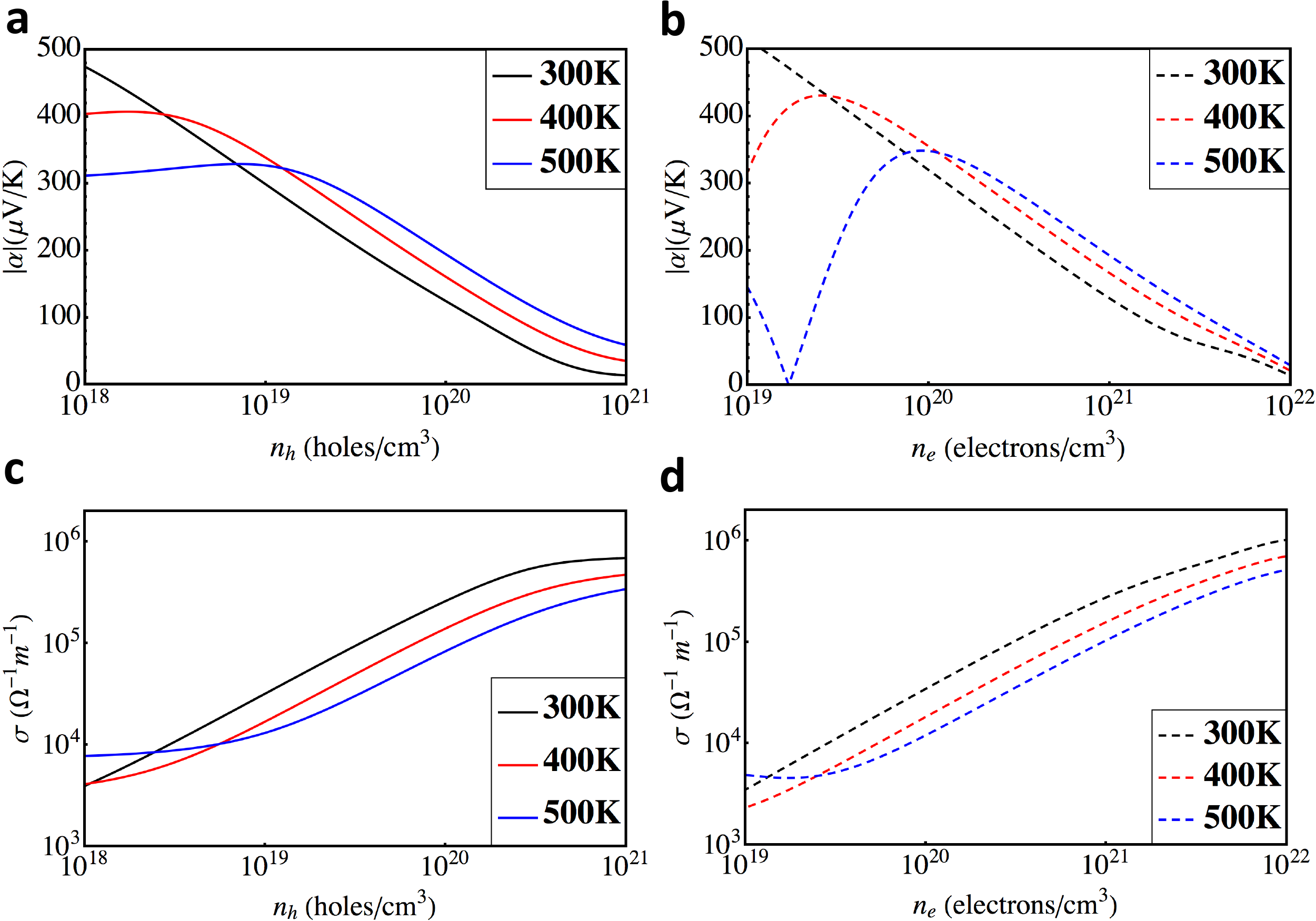}
\caption{The \textbf{a)} $p$-type and \textbf{b)} $n$-type absolute Seebeck coefficients. The \textbf{c)} $p$-type and \textbf{d)} $n$-type conductivities. Polar optical scattering is included.} 
\label{fig:fe2tisitransport}
\end{figure}

A few notable observations are to be made at this stage. With full polar optical scattering, the $p$-type and $n$-type Seebeck coefficients are nearly equivalent at the respective optimal carrier concentrations of $n_{h}\sim10^{20}$ cm$^{-3}$ and $n_{e}\sim10^{21}$ cm$^{-3}$ (see Fig. \ref{fig:fe2tisitransport}). At 500 K, for instance, the optimal $n$-type $\alpha$ is larger only by $3$ $\mu$V K$^{-1}$, which only contributes a 1\% increase in the $n$-type PF over the $p$-type PF, a negligible amount. Meanwhile, the optimal $n$-type $\sigma$ is higher than than the optimal $p$-type $\sigma$ by a factor of nearly 1.18, which leads to an 18\% increase in the $n$-type PF over the $p$-type PF. Evidently, conductivity, not the Seebeck coefficient fuels the comparative $n$-type PF. Since lifetimes are clearly lower for the conduction bands and the valence bands are arguably more dispersive, this phenomenon can only be explained by the presence of an extra fourth and fifth dispersive branches at the CBM over three at the VBM. 

These observations lead to the conclusion that, relative to the $p$-type case, the $n$-type PF draws benefits not from the flat-and-dispersive nature of the bands but rather from the higher number of dispersive pockets. Conversely stated, the $p$-type PF is on par with the $n$-type PF in spite of \text{and} only because of the disadvantage in dispersive pocket multiplicity. If additional dispersive pocket(s) were present, purely dispersive bands could easily deliver higher conductivity and higher PF than flat-and-dispersive bands. A higher dispersion likewise would lead to even better thermoelectic properties via longer lifetime, higher group velocity, with little difference in the Seebeck coefficient.

As a side-note, Fe$_{2}$TiSi would not make a useful thermoelectric as either $p$-type or $n$-type due to intrinsically high lattice thermal conductivity ($> 50$ W m$^{-1}$ K$^{-1}$). Our analysis suggests that its $zT$ would be less than 0.5 even with generous 4\% Hf-Ti substitution that could lower $\kappa_{\text{lat}}$ to $5$ W m$^{-1}$ K$^{-1}$.

\bibliography{references}